# Automated folding of origami lattices: from nanopatterned sheets to stiff meta-biomaterials

Teunis van Manen[1*], Mahya Ganjian[1], Khashayar Modaresifar[1], Lidy E. Fratila-Apachitei[1] and Amir A. Zadpoor[1]

[1]*Department of Biomechanical Engineering, Delft University of Technology (TU Delft), Mekelweg 2, Delft 2628CD, The Netherlands.*
[*]*Email: t.vanmanen@tudelft.nl*

**ABSTRACT**
Folding nanopatterned flat sheets into complex 3D structures enables the fabrication of meta-biomaterials that combine a rationally designed 3D architecture (*e.g.*, to tune mechanical and mass transport properties) with nanoscale surface features (*e.g.*, to guide the differentiation of stem cells). Self-folding is an attractive approach for realizing such materials. However, self-folded lattices are generally too compliant as there is an inherent competition between the ease-of-folding requirements and the final load-bearing characteristics. Inspired by sheet metal forming, we propose an alternative route for the fabrication of origami lattices. This 'automated folding' approach allows for the introduction of sharp folds into thick metal sheets, thereby enhancing their stiffness. We then demonstrate the first time ever realization of automatically folded origami lattices with bone-mimicking mechanical properties (elastic modulus = 0.5 GPa). The proposed approach is highly scalable given that the unit cells making up the meta-biomaterial can be arbitrarily large in number and small in dimensions. To demonstrate the scalability and versatility of the proposed approach, we fabricated origami lattices with > 100 unit cells, lattices with unit cells as small as 1.25 mm, and auxetic lattices. We then used inductively coupled plasma reactive ion etching to nanopattern the surface of the sheets prior to folding. Protected by a thin layer of coating, these nanoscale features remained intact during the folding process. A cell culture assay was then used to assess the bone tissue regeneration performance of the folded origami. We found that the nanopatterned folded specimens exhibit significantly increased mineralization as compared to their non-patterned counterparts.
**Keywords:** Designer biomaterials; origami; biofunctionalization; foldable medical devices.

## 1. INTRODUCTION

Combining self-folding with nanopatterning allows for the fabrication of multi-functional meta-biomaterials blending the advantages of a fully-accessible flat surface with those of a rationally designed 3D architecture [1-3]. The precisely controlled nanopatterns can, for example, kill bacteria [4, 5] and cause the osteogenic differentiation of stem cells [6], while the 3D porous architecture can adjust the mechanical [7-9] and mass transport [10, 11] properties of the obtained lattices while also affecting the curvature sensed by the cells [12]. The incorporation of multiple functionalities through the folding of flat constructs has applications in other areas as well, including flexible electronics [3, 13] and soft robotics [14].

Given the numerous advantages of such origami lattices, many dedicated self-folding techniques have been developed in the recent years [15-17]. Nevertheless, several major challenges remain unresolved. Most importantly, self-folded origami lattices are usually too compliant to be suitable for such applications as load-bearing bone substitutes. Second, it remains highly challenging to self-fold origami lattices whose unit cells are large in number (*i.e.*, hundreds to thousands) and small in dimensions (*i.e.*, meso/microscale range). The second challenge can, in principle, be addressed within the paradigm of self-folding materials, for example, by developing 4D printing techniques with finer resolutions [18, 19] and synthesizing shape memory polymers with higher magnitudes of generated forces [20]. The first challenge, however, is inherent to self-folding processes. That is because there is an inherent competition between the ease-of-folding requirements on the one hand and the final load-bearing characteristics of self-folded lattices on the other. Indeed, we have recently demonstrated that there exist theoretical boundaries limiting the stiffness of self-folded origami [21].

One approach to address the first limitation is to use an additional (locking) mechanism to enhance the mechanical properties of the folding elements after completing the self-folding process. For example, separate locking elements are used in a few studies to increase the stiffness of the folded structure [22-24]. Another method includes the so-called 'layer-jamming mechanisms' in which a stack of self-folding elements is compressed after folding to enhance their combined stiffness [25, 26]. However, both techniques highly limit the design freedom and scalability.

Here, we propose the alternative approach of 'automated folding' to address both abovementioned limitations of self-folding processes. Inspired by sheet metal forming processes, we developed miniaturized folding devices that can fold origami lattices made from a large number of small unit cells. This approach allows for realizing sharp permanent folds in thick (metallic) sheets. The sharpness of the folds, the large thickness of the sheets, and the freedom in using any material including metals translates to high, bone-mimicking stiffness values while also enabling the inclusion of pore sizes that are suitable for bone tissue regeneration (*i.e.*, 300-900 µm [27-29]). The afforded scalability in the design of meta-biomaterials also extends to the dimensions of the unit cells, as the proposed approach is highly amenable to miniaturization. Indeed, the primary factor limiting the unit cell dimensions is the precision of the applied micromachining technique. We use the proposed approach to demonstrate the first ever realization of automatically folded origami lattices with bone-mimicking mechanical properties. To showcase the versatility of the unit cells that can be realized, we also fold origami lattices with auxetic properties (*i.e.*, a negative Poisson's ratio). Finally, we demonstrate the entire process of fabricating origami-based meta-biomaterials by micromachining thick titanium sheets followed by surface nanopatterning using inductively coupled plasma reactive ion etching (ICP RIE), application of a protective layer to preserve the nanopatterns during the folding process, and the biological evaluation of the folded specimens using *in vitro* cell culture assays.

## 2. RESULTS AND DISCUSSION

### 2.1. Automated folding of origami lattices

We start by considering regular lattices made of an identical unit cell repeated along the three orthogonal axes (Figure 1a). Such lattices can be unfolded by slicing their 3D geometry into 'storeys' whose thickness equals that of a single unit cell (Figure 1a). Depending on the type of the unit cell, different folding patterns have been proposed in the literature for the folding of such storeys [2, 30]. Here, we will be focusing on one such folding strategy (Figure 1b) while keeping in mind that the other folding strategies can, in principle, be realized as well. First, a series of collinear hinges along the y-direction are simultaneously folded into a corrugated sheet, in a row-by-row manner (Figure 1b). Upon the completion of the first stage, all the remaining (*i.e.*, unfolded) crease lines align along the z-direction. Second, the corrugated sheet is folded into the final array of cubes (Figure 1b). This step can be repeated sequentially to achieve the desired array of the unit cells. The main advantage of this particular folding sequence is that the folding steps can be decoupled into multiple non-interfering stages. Consequently, a simple, yet highly scalable folding device can be designed to automate the folding process. In addition, the selected folding pattern is rigid-foldable, meaning that the deformation is localized at the crease lines. The faces are, therefore, not stretched or twisted during the folding process, ensuring that the applied surface ornaments remain intact. Finally, the lattices can be folded from a single sheet, further improving their structural integrity.

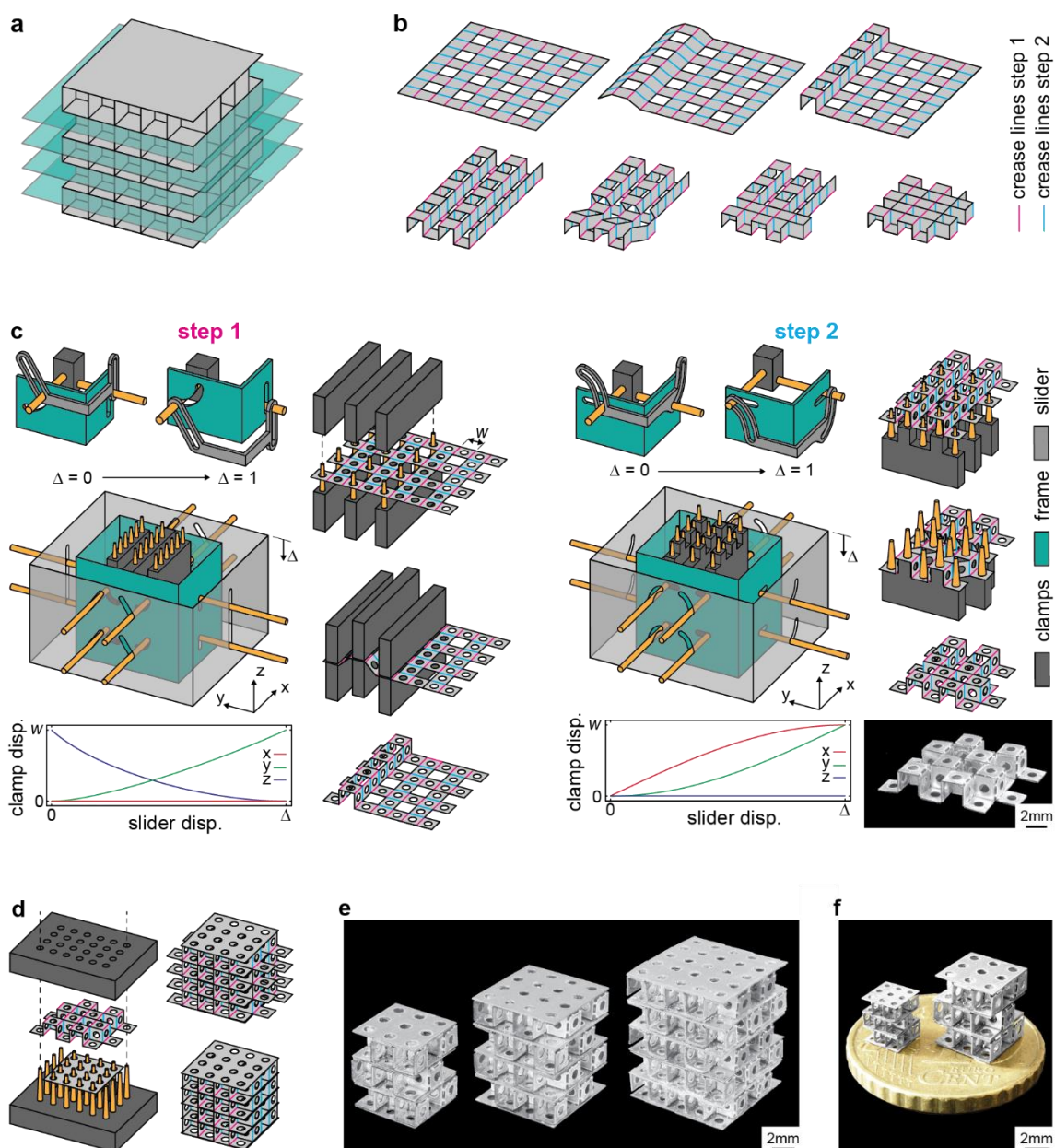

**Figure 1. Fabrication of stiff cubic lattices.** (a) The 'slicing' of a cubic lattice structure. (b) The folding sequences of an array of cubic unit cells. (c) A schematic illustration of the two-step folding process of cubic 'storeys' using external folding devices. (d) The assembly of the folded 'storeys' into a cubic lattice structure. (e) The photographs of the folded cubic lattices with different numbers of unit cells. (f) The cubic lattices with different unit cell dimensions ($w$ = 1.25 mm and $w$ = 2.0 mm) are shown on top of a 10 Euro cent coin.

We designed two sets of folding devices (*i.e.*, one for each folding step, Figure 1c). Stereolithography (SLA) 3D printing was used for the fabrication of the folding devices (see Materials and Methods, Figure S1 in the supplementary document). First, we designed clamps that use small pins to hold the flat sheets. Circular holes were, therefore, incorporated into the design of the initially flat sheets. The clamps can be actuated using a mechanism that is made of a frame with fixed slots and an outer slider containing a number of guiding slots (Figure 1c). The frame and sliders are linked through the pins connected to the two moving clamps. The middle clamp is fixed to the frame. Only actuated by the downward displacement of the slider, $\Delta$, the clamps make the desired movement that is encoded into the geometry of the slots (Figure 1c). In the first folding step, the specimen is clamped between two similar folding devices to constrain the displacement of the specimen in the z-direction. For the second folding step, no constraints in the z-direction need to be applied and a single folding device is sufficient for the folding of the specimen. Because the folding sequence is composed of two independent steps, large arrays of polyhedra can be folded provided that the widths of the folding devices are large enough.

Following the folding process, the individual 'storeys' need to be assembled into a cubic lattice. Additional flat sheets were designed that provided the surface area for the adhesive bonding of the folded 'storeys' (Figure 1d). A special device made of an array of pins and holes was used to align the folded specimens and to press them together during the curing time of the adhesive. The assembled cubic lattices still contain a few protruding panels that were needed for the clamping of the specimen during the folding process. Upon the manual removal of these panels, the final cubic lattice is obtained (Figure 1d). Multiple cubic lattices with different numbers of unit cells ($w$ = 2.0 mm) were folded and assembled using the proposed approach (Figure 1e). In addition, smaller folding devices were used for the fabrication of a cubic lattice structure with $w$ = 1.25 mm (Figure 1f and Figure S4), demonstrating the scalability of the developed folding technique.

**2.2. Mechanical properties of cubic lattices**
Given the importance of mechanical properties in determining the performance of load-bearing meta-biomaterials, we studied how origami lattices respond to mechanical loads. Initially flat sheets with a panel width ($w$) of 2.0 mm were laser-cut from titanium foils with a thickness of 125 µm. Crease lines were integrated into the design in order to concentrate the deformation at the folding lines (Figure 2a). The length of one of the panels was slightly extended (Figure 2a) such that the edges of this panel aligned with the top and bottom panels and could, thus, contribute to the load-bearing capacity of the specimens when loaded in compression. Circular holes were incorporated into the design to not only serve as the clamping sites for the folding devices but to also produce an interconnected array of cubic unit cells. In general, meta-biomaterials need to be permeable to enable both cell migration and the transport of oxygen and nutrients [31]. Scanning electron microscope (SEM) images were used to assess the quality of the specimens after folding. We found that the deformations were localized at the crease lines without any signs of cracks (Figure 2b).

For the assembly of the folded lattices, a biocompatible cyanoacrylate-based adhesive was selected. The mechanical performance of the adhesive was assessed by performing a series of lap shear tests (Figure 2c). The results reveal an average shear strength of 14.1 MPa and an average shear modulus of 270 MPa, which are in the same range as other cyanoacrylate-based adhesives [32-34] and are expected to be sufficient for carrying the shear loads developed within the storeys due to the potential misalignments and the other unforeseen causes of shear loading.

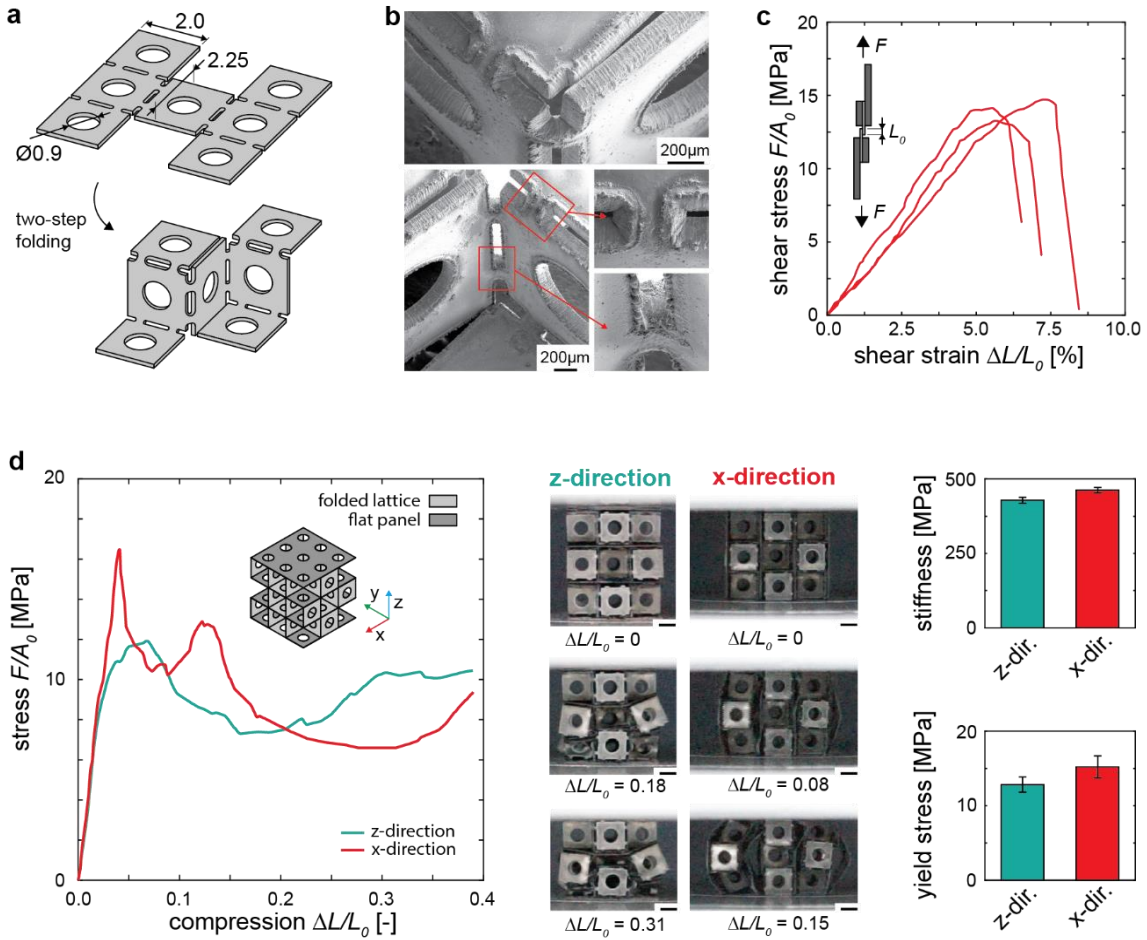

**Figure 2. The mechanical properties of the folded cubic lattices.** (a) A schematic representation of the initial geometry of the flat material and the folded structure. (b) The SEM images of a folded cubic structure showing the sharp folding lines. (c) The lab shear test results for the biocompatible adhesive ($n = 3$). (d) The compression test results of the folded lattices containing 3×3×3 cubic unit cells ($n = 3$).

Subsequently, compression tests were performed on the origami lattices made from an array of 3×3×3 cubic unit cells. For the first set of experiments, the specimens were oriented along the z-direction. The specimens failed though the sequential collapse of individual 'storeys' (Figure 2d). We also compressed the specimens along the x-direction and observed that the dominant failure mode shifts to the failure of the adhesive accompanied by the outward buckling of the outer storeys (Figure 2d). Regardless of the loading direction, the effective elastic modulus was 400-500 MPa while the yield strength was around 15 MPa (Figure 2d). These values are in the same range as the mechanical properties of the cancellous bone [35] and confirm the bone-mimicking mechanical properties of the folded origami lattices. Samples containing more unit cells are expected to exhibit even higher stiffness values because the panels located on the edges of the specimens carry less loads (Figure 1e). Confining the folded lattices within the boundaries of a cavity may also increase their effective stiffness, as outward buckling would be suppressed. Finally, the design of the crease lines can be further optimized to increase both the stiffness and the buckling load of individual storeys.

## 2.3. Auxetic and non-auxetic hexagonal lattices

In addition to cubic lattices, several other types of origami can be automatically folded using the proposed approach. Given the importance of auxetic structures in the design of meta-biomaterials [36], we folded origami lattices based on the re-entrant unit cell. Similar to the previous designs, the storeys were folded in two independent steps (Figure 3a), meaning that the automated folding method is equally scalable and the maximum number of unit cells is only limited by the width of the folding device. Figure S2 (supplementary document) presents additional details regarding the design of the folding devices. From the same initially flat sheet, specimens with re-entrant angles ($\theta$) of up to 105° can be folded using a single set of folding devices. The re-entrant angle can be controlled depending on the amount of the downward displacement of the slider, $\Delta$. As a demonstration, two specimens were fabricated with $\theta$ angles equal to 75° and 105° (Figure 3b). Both specimens were compressed to measure their Poisson's ratios. Image processing was used to determine how the mean value of the Poisson's ratio evolves as the compressive load gradually increases. Indeed, the specimen with a $\theta$ angle equal to 75° exhibited a positive Poisson's ratio ranging between 0.5 and 1.5 depending on the magnitude of the applied strain (Figure 3c). A negative Poisson's ratio between -2.0 and -1.0 was observed for the sample with a $\theta$ angle equal to 105° (Figure 3d). These results clearly show the potential of the proposed approach for folding origami lattices based on various other types of unit cells. The ability to adjust the type of the unit cell means that automatically folded meta-biomaterials can offer a wide range of (mechanical) properties and functionalities.

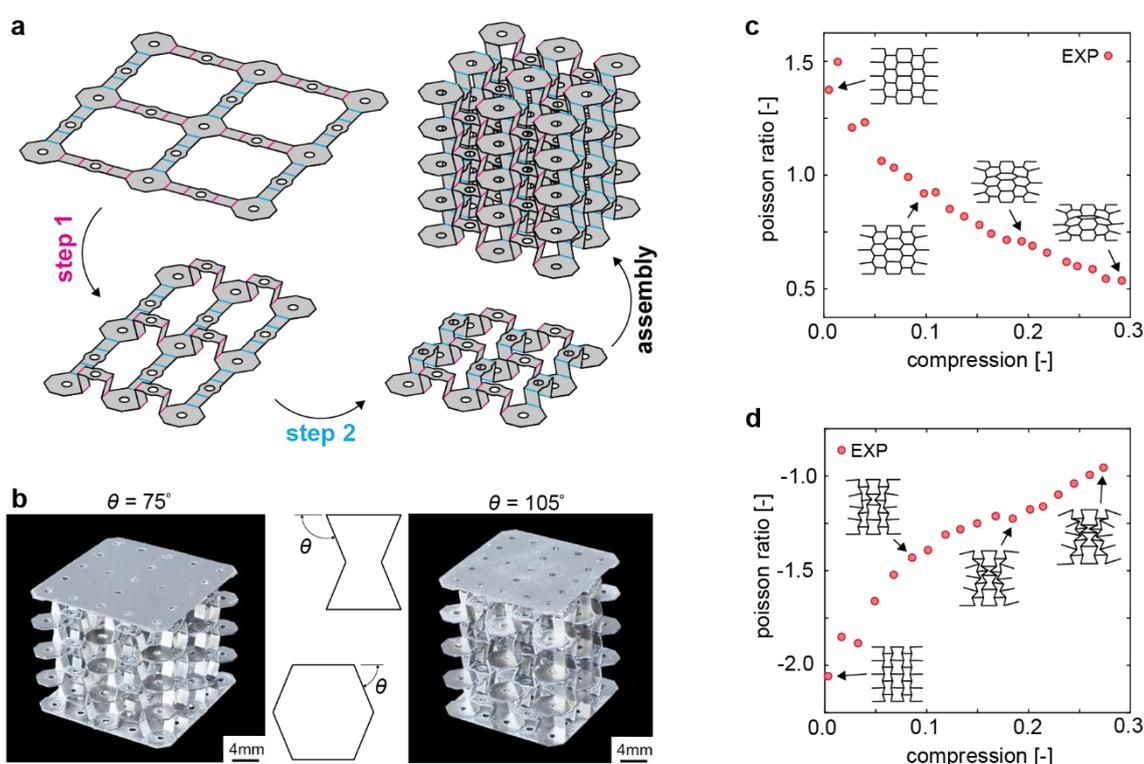

**Figure 3. Folded re-entrant structures.** (a) A schematic representation of the two-step folding process and the subsequent assembly of the re-entrant lattice structures. (b) The photographs of the fabricated auxetic lattices with different values of the re-entrant angle, $\theta$. The Poisson's ratio as a function of the compressive strain for lattices with $\theta$ angles of (c) 75° and (d) 105°.

## 2.4. Nanopatterning

The main advantage of the lattices folded from initially flat sheets is the ability to functionalize the entire surface of the final implant using advanced nanopatterning techniques. During the folding process, the functionalized surface needs to be protected to prevent it from being damaged by the folding tools and the self-contact of the sheet. We, therefore, applied a thin layer of polyvinyl alcohol (PVA) coating to the nanopatterned specimens [37]. Upon the completion of the folding process, the coating was dissolved by immersing the specimens in hot water (Figure 4a). We used dry etching (ICP RIE) to pattern the surface of the polished flat specimens with nanopillars and create the so-called black titanium (bTi) specimens. The parameters of the ICP RIE were selected such that the resulting nanopillars had dimensions in the range known to be both osteogenic and bactericidal [38]. Evaluating the generated bTi nanopillars by SEM and AFM (data not shown) revealed that they had a tip diameter below 100 nm and a height between 700 nm and 1000 nm. These dimensions have been previously shown to induce both osteogenic and bactericidal properties [38]. After each production step, SEM images were made of the nanopatterned samples to determine whether the applied nanostructures were still intact. These images confirmed that the patterns remain unaffected by the different production steps (Figure 4b), save for a small laser heat-affected zone (Figure 2b).

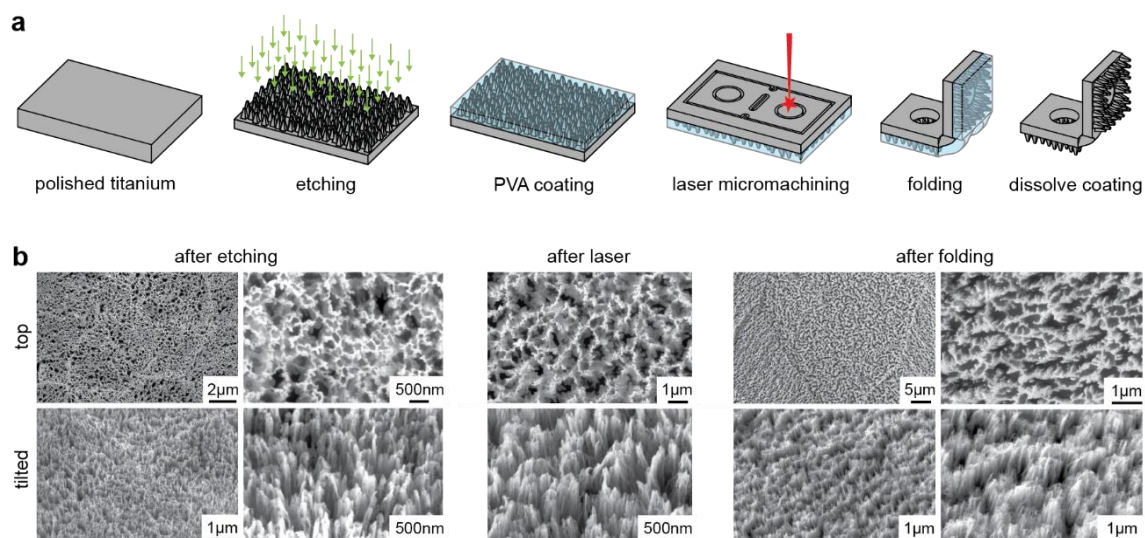

**Figure 4. Surface functionalization.** (a) A schematic illustration of the nanopatterning and surface protection steps. (b) The SEM images of the bTi surface at different stages of the production process.

## 2.5. Cell culture

We performed biological assays to compare the osteogenic properties of the nanopatterned lattices with their non-patterned polished counterparts. Both patterned and non-patterned specimens supported the attachment of the preosteoblasts onto their surface and did not hinder the expression of Runx2 as an osteogenic marker (Figure 5a). There was no significant difference in the intensity of the Runx2 signal between the patterned and non-patterned specimens (Figure 5a). The results obtained here for 3D specimens is in line with the results of a previous *in vitro* study on 2D substrates covered with different configurations of bTi pillars [38] in which we found that a configuration of pillars similar to the present study did not downregulate the expression of Runx2 while other nanopattern designs (*e.g.*, taller and more separated pillars) significantly impaired it. Furthermore, in agreement with the previous studies [38, 39], the percentage of the area containing mineralized nodules was significantly higher for the nanopatterned origami lattices as compared to the non-patterned specimens (Figure 5b). The enhanced mineralization induced by bTi nanopillars has been argued to be associated with the early stage changes in the adaptation of the cells to the surface, including the changes in the formation and distribution of focal adhesions as well as changes in the cell morphology [38, 39]. The full consistency of the results obtained in the current 3D experiments with those of our previous 2D studies suggest that the functionalities observed in flat specimens carry over to the folded origami lattices. This is an important conclusion that, if supported by further evidence, could greatly simplify the process of evaluating the biological response of origami-based meta-biomaterials. Further studies, including experiments using *in vivo* animal models, are required to explain the effects of such nanopatterned constructs on the bone tissue regeneration process. Moreover, nanopatterns such as those applied here have the potential to induce additional functionalities, such as bactericidal properties [38]. Combining an osteogenic response with bactericidal properties would enable the origami-based meta-biomaterials to both improve bony ingrowth and prevent implant-associated infections.

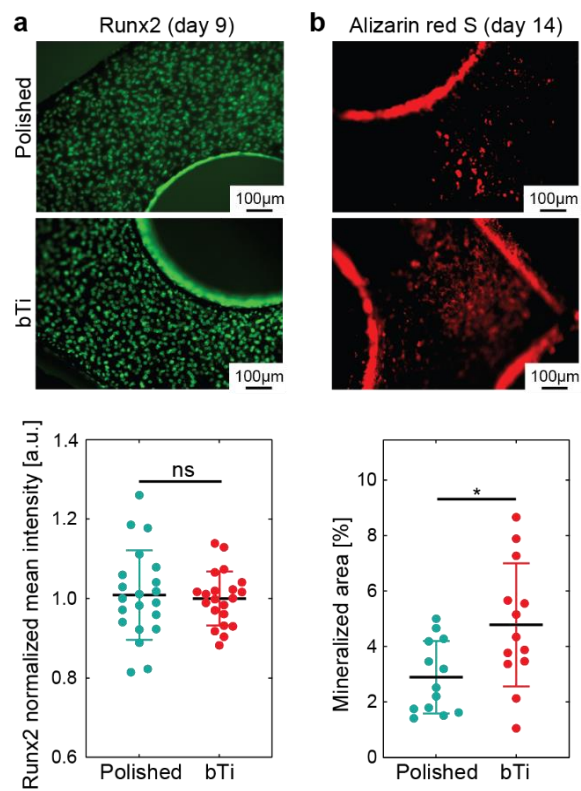

**Figure 5. The osteogenic response of the preosteoblast cells cultured on the folded lattice structures.** (a) The representative immunofluorescent images of Runx2 after 9 days of culture. No significant difference between the polished and bTi lattices was found. (b) The representative images of the mineralized nodules after Alizarin red S staining after (day 14). The mineralized area is significantly larger for the bTi lattices as compared with the polished specimens. Unpaired *t*-test with the Welch correction, *$p < 0.05$.

## 3. OUTLOOK AND CONCLUSIONS

In summary, we presented a methodology for the fabrication of stiff origami lattices whose (internal) surfaces were decorated with nanoscale surface ornaments. While this method could, in principle, be applied to a wide range of materials, we demonstrated automatically folded cubic lattices with an effective modulus of up to 0.5 GPa, automatically folded auxetic and non-auxetic hexagonal lattices, and folded lattices with osteogenic surface nanopatterns. The application of a thin coating layer was sufficient to protect the surface nanopatterns during the laser cutting and subsequent folding steps. Finally, our *in vitro* assays confirmed that the nanopatterned origami lattices exhibit enhanced osteogenic response as compared to their non-patterned counterparts, including a significantly higher degree of mineralization.

Bone-substituting meta-biomaterials require a pore size within the range 200-2000 μm [40, 41], although the optimum pore size is generally believed to be between 300-900 μm [27-29]. To demonstrate the scalability of the presented folding strategy, we folded a cubic lattice with a unit cell size of 1250 μm from a 50 μm thick titanium sheet (Figure 1f). In principle, further miniaturization would be possible provided that more advanced production techniques, such as micro-milling [42, 43], are used for the fabrication of the folding devices.

It is generally assumed that, to avoid stress shielding and stimulate bone tissue regeneration, the mechanical properties of bone substitutes should match those of their surrounding bone tissue [31]. Dependent on the type and location of the bony defect, an elastic modulus of 0.1-20 GPa may be required [44, 45]. Here, we reported the fabrication of lattice structures with a maximum effective modulus of 0.5 GPa, which is in the range of the values reported in the literature for the trabecular bone [35]. One method to increase the elastic modulus of the folded structures is to reduce the size of the unit-cells meaning that more unit-cells fit in a similar volume, thereby increasing the load-bearing capacity of the structure [21]. An alternative strategy would be the stacking of a number of loosely connected sheets with relatively smaller thicknesses that are simultaneously folded to achieve the desired combination of large thickness values and an increased bend sharpness [21].

We only applied the surface nanopatterns to one side of the sheets used here. Applying the nanopatterns to both sides of the sheets is expected to further enhance the osteogenic potential of the resulting meta-biomaterials. While there is no fundamental reason why this may not be possible, there are a few practical challenges that need to be overcome first. In particular, bespoke techniques need to be developed to protect the already patterned side while the other side is being patterned.

Given that stiff biofunctionalized origami lattices comprising large arrays of small interconnected unit-cells can be fabricated, the described approach paves the way for the application of folded lattices as load-bearing meta-biomaterials. Such metamaterials can be fabricated from a wide range of materials while the different types of surface-related functionalities could be embedded into the initially flat material. Even though some aspects of the applied methodology could benefit from further optimization, the proposed approach satisfies all the primary design objectives of a meta-biomaterial, including biocompatibility, fully customizable surface nanopatterns, bone-mimicking mechanical properties, osteogenic behavior, appropriate range of pore sizes, and scalability. The presented production technique could, therefore, serve as a platform for further biological evaluation of origami-inspired meta-biomaterials, including both *in vitro* and *in vivo* experiments.

## 4. MATERIALS AND METHODS

### 4.1. Folding devices
Folding devices were manufactured using a stereolithography (SLA) 3D printer (Form 3, Formlabs, Massachusetts, US). CNC milling was used for the precise drilling of the holes in the frame and the clamps for the clamping pins. Finally, the clamps, frame, and slider were assembled using steel rods (diameter = 2.0 mm).

### 4.2. Sample fabrication
The specimens were produced from pure titanium sheets (titanium foil, purity = 99.6+%, annealed, Goodfellow, UK) with a thickness of either 50 µm or 125 µm. The specimens were laser cut using laser micromachining equipment (Optec Laser Micromachining Systems, Belgium). A biocompatible cyanoacrylate-based adhesive (ethyl cyanoacrylate, MB297Med-2, MasterBond, US) was used for to assemble the folded storeys. After the application of the adhesive, the specimens were compressed for at least 2 minutes to guarantee the sufficient fixation of the components. The mechanical tests were performed after at least 48 h to ensure the adhesive was fully cured. Figure S3 (supplementary document) illustrates the various steps of the production process.

### 4.3. Mechanical testing

A Lloyd-LR5K mechanical testing machine equipped with a 100 N load-cell was used for the compressive mechanical testing of the folded specimens. The resulting deformations were captured using a high-resolution digital camera (Sony A7R, lens = Sony FE 90-mm f/2.8 macro OSS). Lines were fit to the experimental data using a custom MATLAB (Mathworks, version R2020b, US) code to determine the stiffness of the folded lattices. Custom MATLAB codes were used for image processing in order to obtain the Poisson's ratio of the lattices made using the re-entrant unit cells (Figure 3).

### 4.4. Nanopatterning
Annealed titanium sheets (titanium foil, purity = 99.6+%, annealed; Goodfellow, UK) with a thickness of 125 μm were used for the fabrication of the bTi samples. The sheets were cut to the size of a 4-inch (diameter =101.6mm) silicon wafer and were polished by chemical mechanical polishing (CMP Mecapol E460, Saint-Martin-le-Vinoux, France). The 4-inch titanium sheets were then cut into $10 \times 14$ mm$^2$ pieces using the laser micromachine. Consequently, the samples were cleaned in acetone, ethanol, and isopropyl alcohol (IPA) (consecutive steps of 30 min each), and were spin-dried.

The polished Ti specimens were nanopatterned using an ICP RIE machine (PlasmaLab System 100, Oxford Instruments, UK) to create the so-called black titanium (bTi) specimens. First, the Ti specimens were glued with a thermal joint compound (type 120, Wakefield Engineering, USA) to a 4-inch fused silica carrier wafer. The etching gasses included $Cl_2$ and Ar. The etching process was performed under the following condition: ICP power = 600 W, RF power = 100 W, $Cl_2$ flow rate = 30 sccm, Ar flow rate = 2.5 sccm, chamber pressure = 0.02 mbar, temperature = 40 °C, and etching time = 10 minutes. Following the etching process, the specimens were cleaned in acetone, ethanol, and IPA (consecutive steps of 30 min each), and were then spin-dried. High-resolution scanning electron microscopy (SEM) images were taken with a Helios (FEI Helios G4 CX dual-beam workstation, Hillsborough, OR, USA) microscope at 10 kV and 43 pA. SEM and atomic force microscopy (AFM, JPK Nanowizard 4, Bruker, Germany) were used to measure the height and tip diameter of the nanopillars as described before [38]. To prevent the bTi nanopatterns from being damaged during the following laser-cut and folding processes, the specimens were coated two times with polyvinyl alcohol (PVA) (Resin Technology, France), each step followed by thermal treatment at 50 °C for 1 hour. This layer was removed prior to the cell culture experiments by placing the folded specimens in hot water (temperature = 70 °C) overnight.

### 4.5. Cell culture
Prior to cell seeding, all the fabricated specimens were sterilized through immersion in 70% ethanol and exposure to UV light for 20 minutes. Preosteoblast MC3T3-E1 cells (Sigma-Aldrich, Germany) were cultured in alpha minimum essential medium (α-MEM) supplemented with 10% (v/v) fetal bovine serum (FBS) and 1% (v/v) penicillin-streptomycin (all from Thermo Fisher Scientific, US). After reaching confluence, the cells were seeded on both types of constructs ($n = 4$) with a density of $2 \times 10^4$ cells per construct in a 48 well-plate, and were incubated at 37 °C and 5% $CO_2$. The cell culture medium was refreshed every 2 days by adding 50 μg/ml ascorbic acid (1:1000) and 4 mM β-glycerophosphate (1:500) (both from Sigma-Aldrich, Germany) to α-MEM. After 9 days of culture, the cells adhered to the constructs were washed with phosphate buffered saline (PBS) and were fixated using a 4% (v/v) formaldehyde solution (Sigma-Aldrich, Germany). The cell membrane was permeabilized by adding 0.5% Triton X-100/PBS (Sigma-Aldrich, Germany) at 4 °C for 5 min. Subsequently, the constructs were incubated in 1% BSA/PBS (Sigma-Aldrich, Germany) at 37 °C for 5 min. Then, the cells were incubated in recombinant anti-Runx2 rabbit monoclonal primary antibody (1:250 in 1%

BSA/PBS, Abcam, UK) for 1 h at 37 °C. Following three times washing with 0.5% Tween-20/PBS (Sigma-Aldrich, Germany), the cells were incubated in Alexa Fluor 488, donkey anti-rabbit polyclonal secondary antibody (1:200 in 1% BSA/PBS, Thermo Fisher Scientific, US) for 1 h at room temperature. The constructs were finally washed three times with 0.5% Tween-20/PBS and once with 1X PBS. Fluorescent images were acquired from different panels of the constructs using a fluorescence microscope (ZOE™ fluorescent cell imager, Bio-Rad, The Netherlands). Ten images from each study group (*i.e.*, patterned and non-patterned constructs) were randomly selected for further analysis. Three regions of interest (ROIs) with an area of $500 \times 500$ µm$^2$ were analyzed for the intensity of Runx2 using ImageJ 1.53c (NIH, US). Finally, the mean intensity of each ROI was normalized with respect to the mean intensity of all ROIs within the same image.

To analyze the matrix mineralization induced by the constructs, the Alizarin red S assay was performed after 14 days of culture. The cells were washed and fixated as described above. The constructs were then incubated in 2% (w/v) Alizarin red S solution (Sigma-Aldrich, Germany) for 30 min in the dark. The constructs were then rinsed 5 times with distilled water before being imaged by a ZOE™ fluorescent cell imager (Bio-Rad, The Netherlands). The total area of the mineralized nodules on the surface was quantified by running the Analyze Particles command in ImageJ then normalized with respect to the area of the surface on which the nodules were observed.

**ACKNOWLEDGEMENTS**
The research leading to these results has received funding from the European Research Council under the ERC grant agreement no. [677575].

# Automated folding of origami lattices: from nanopatterned sheets to stiff meta-biomaterials

Teunis van Manen[1*], Mahya Ganjian[1], Khashayar Modaresifar[1], Lidy E. Fratila-Apachitei[1] and Amir A. Zadpoor[1]

[1]*Department of Biomechanical Engineering, Delft University of Technology (TU Delft), Mekelweg 2, Delft 2628CD, The Netherlands.*
*Email: t.vanmanen@tudelft.nl

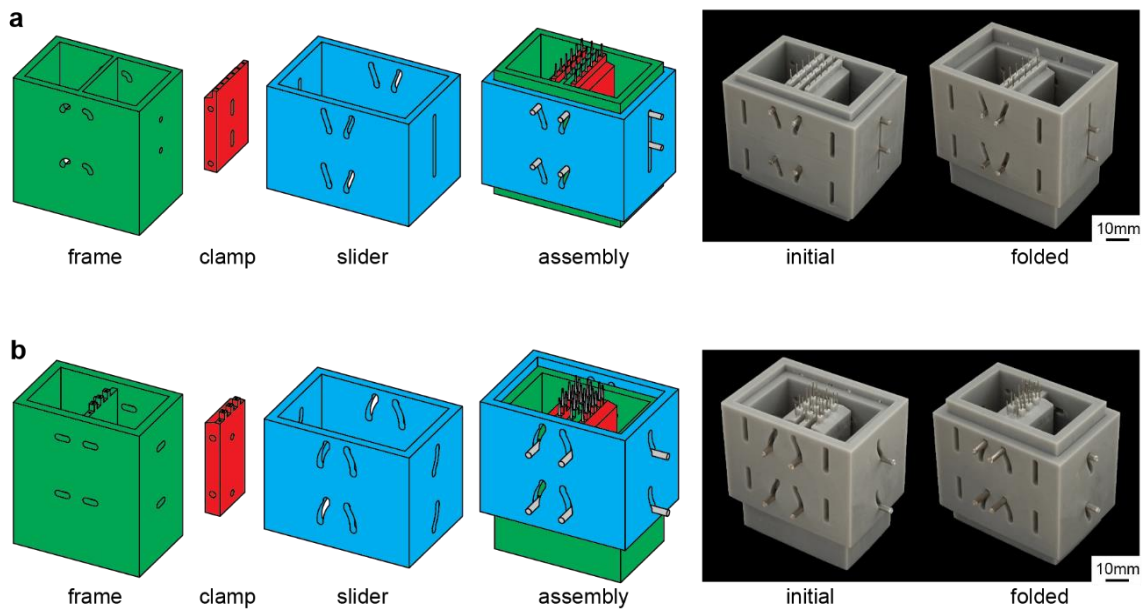

**Figure S1. Folding devices for the folding of cubic lattice structures.** Schematic representations and pictures of the folding devices used for the folding of cubic lattices for folding step 1 (a) and folding step 2 (b).

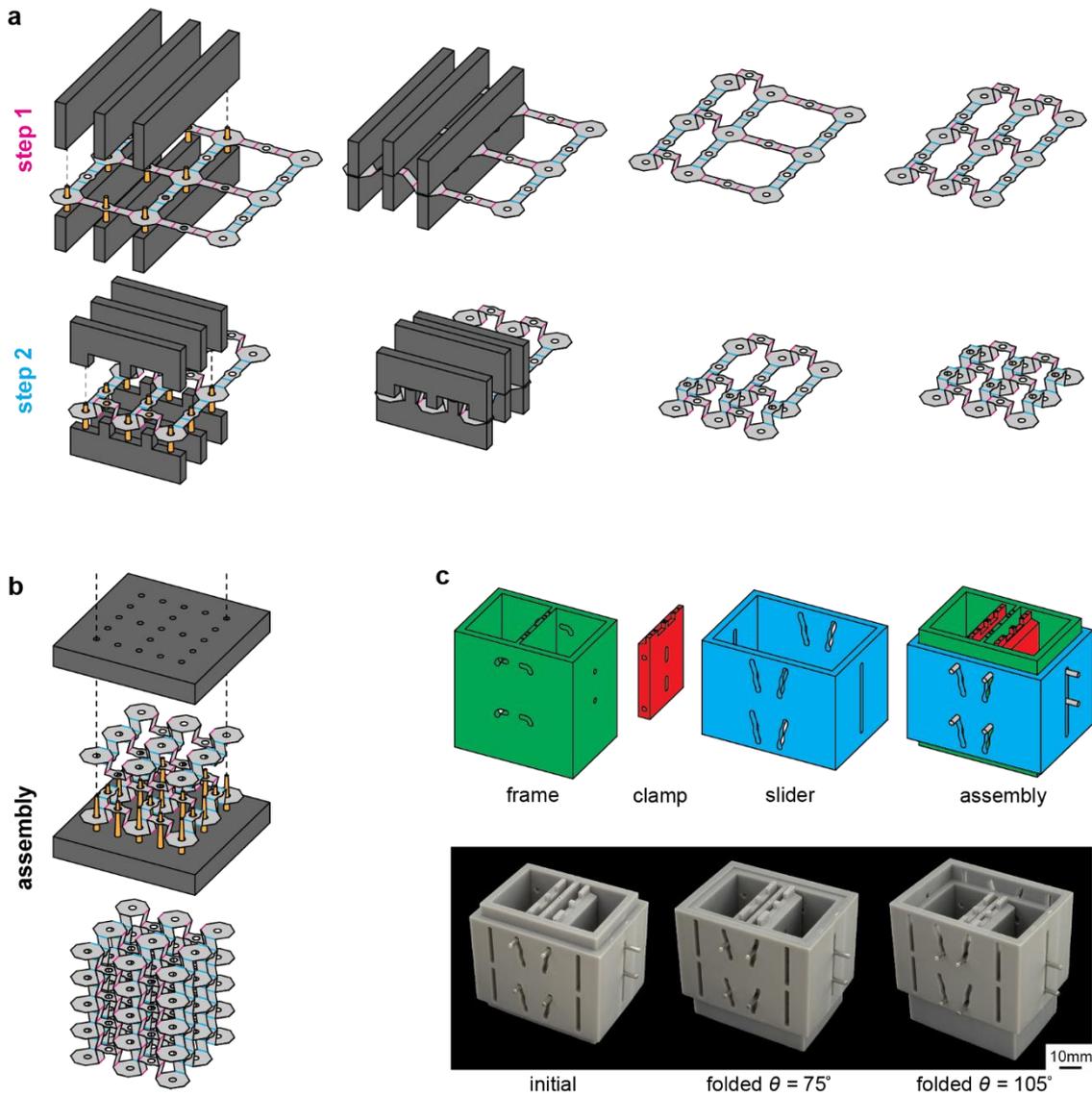

**Figure S2. Folding of re-entrant structures.** (a) Schematic illustration of the folding of re-entrant 'storeys'. (b) Assembly of the folded structures into a re-entrant lattice. (c) Schematic representation and pictures of the folding devices used for the folding re-entrant lattice structures. The same folding device can be used for both folding steps. The re-entrant angle $\theta$ can be varied based on the slider displacement.

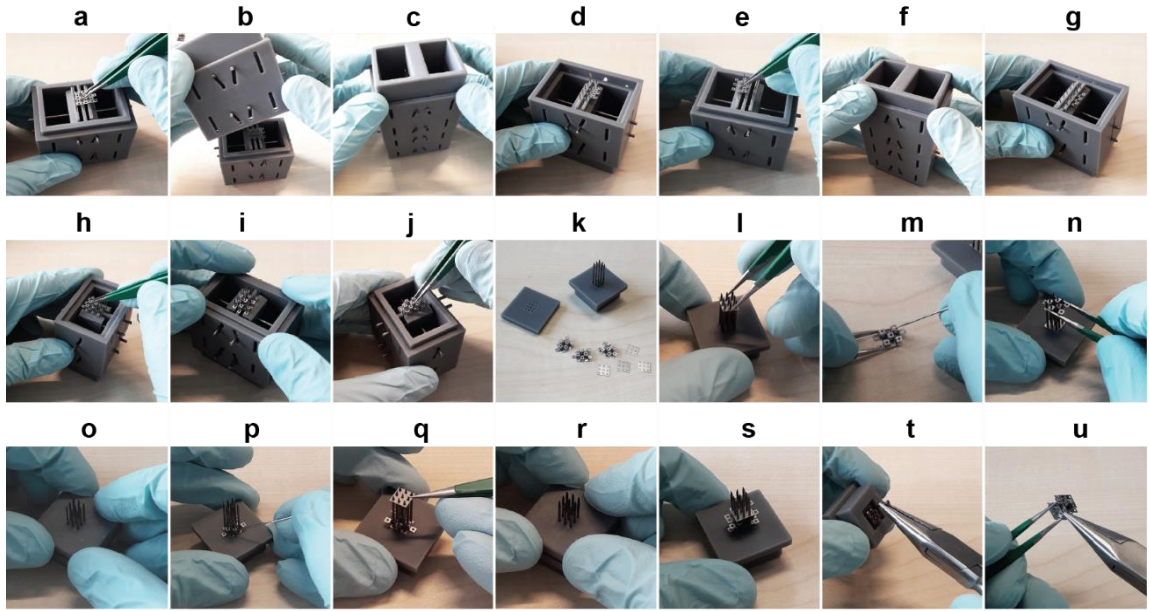

**Figure S3. Pictures showing the production of cubic lattices.** (a-g) The first folding step. (h-j) The second folding step. (k-t) Assembly of the folded 'storeys'. (u) Removal of the protruding panels.

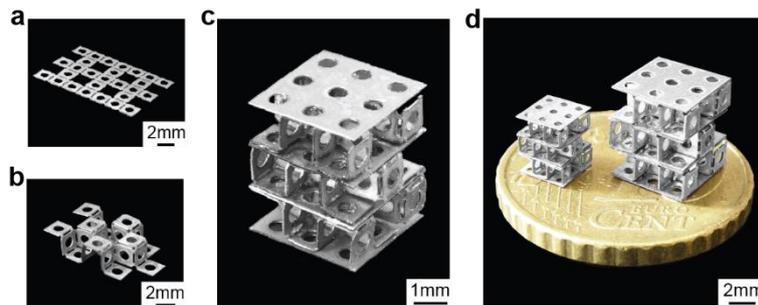

**Figure S4. Fabrication of cubic lattices with unit cell dimensions of 1.25 mm.** (a) Initial flat titanium sheet with thickness of 50 µm. (b) One folded 'storey'. (c) Cubic lattice of 3x3x3 unit cells. (d) Cubic lattices with different unit cell dimensions ($w = 1.25$ mm and $w = 2.0$ mm) are shown side by side with a ten euro cent coin for scale.